\documentclass[twocolumn,tighten,times]{aastex62} %[astrosymb]
\usepackage{newtxmath} %use times font for math

\received{\today}
\revised{? ?, ?}
\accepted{? ?, ?}%\today
\submitjournal{ApJ}

\graphicspath{{./}{figs/}}
\def\simlt{\mathrel{\rlap{\lower 3pt\hbox{$\sim$}}\raise 2.0pt\hbox{$<$}}}
\def\simgt{\mathrel{\rlap{\lower 3pt\hbox{$\sim$}}\raise 2.0pt\hbox{$>$}}}

\shortauthors{Cai et al.}

\begin{document}

\title{High-$z$ dusty star-forming galaxies: a top-heavy initial mass function?}

\correspondingauthor{Zhen-Yi Cai} \email{zcai@ustc.edu.cn}

\author[0000-0002-4223-2198]{Zhen-Yi Cai}
\affiliation{CAS Key Laboratory for Research in Galaxies and Cosmology,
Department  of Astronomy, University of Science and Technology of China, Hefei
230026, China} \affiliation{School of Astronomy and Space Science, University
of Science and Technology of China, Hefei 230026, China}

\author[0000-0003-2868-2595]{Gianfranco De Zotti}
\affiliation{INAF, Osservatorio Astronomico di Padova, Vicolo dell'Osservatorio
5, I-35122 Padova, Italy}

\author[0000-0001-9139-2342]{Matteo Bonato}
\affiliation{INAF, Istituto di Radioastronomia and Italian ALMA Regional
Centre, Via P. Gobetti n. 101, 40129 Bologna, Italy} \affiliation{INAF,
Osservatorio Astronomico di Padova, Vicolo dell'Osservatorio 5, I-35122 Padova,
Italy}

%%%%%%%%%%%%%%%%%%%%%%%%%%%%%%%%%%%%%%%%%%%%%%%%%%%%%%%%%%%%%%%%%%%%%%%%%%%%%%%%
\begin{abstract}
Recent estimates point to abundances of $z>4$ sub-millimeter (sub-mm)
galaxies far above model predictions. The matter is still debated. According
to some analyses the excess may be substantially lower than initially thought
and perhaps accounted for by flux boosting and source blending. However,
there is no general agreement on this conclusion. An excess of $z>6$ dusty
galaxies has also been reported albeit with poor statistics. On the other
hand, evidence of a top-heavy initial mass function (IMF) in high-$z$
starburst galaxies has been reported in the past decades. 
This would translate into a higher sub-mm
luminosity of dusty galaxies at fixed star formation rate, i.e., into a higher
abundance of bright high-$z$ sub-mm galaxies than expected
for a universal Chabrier IMF. Exploiting our physical model for
high-$z$ proto-spheroidal galaxies, we find that part of the excess can be
understood in terms of an IMF somewhat top-heavier than Chabrier.
Such IMF is consistent with that recently proposed to account for the low
$^{13}$C/$^{18}$O abundance ratio in four dusty starburst galaxies at $z=2$--3.
However, extreme top-heavy IMFs are inconsistent with the sub-mm counts at $z > 4$.

\end{abstract}


\section{Introduction}

The opening of the sub-millimeter (sub-mm) window on galaxy formation and
evolution, thanks to the $850\,\mu$m surveys with the Submillimeter Common-User
Bolometer Array (SCUBA)  on the James Clerk Maxwell Telescope \citep[JCMT;
e.g.,][]{Smail1997, Hughes1998, Barger1998}, has revolutionized the field.
SCUBA blank-field pointings demonstrated that the most active star formation
phases of high-$z$ galaxies are heavily dust-enshrouded and therefore largely
missed even by the deepest optical surveys \citep[for a review
see][]{Casey2014}. The results of these surveys have strongly shaken up the
leading galaxy formation paradigm of the time which under-predicted the sub-mm
counts by one order of magnitude or more \citep[e.g.,][]{Kaviani2003,
Baugh2005}.

Part of the discrepancy was later shown to be attributable to an overestimate
of the observed counts \citep{Coppin2006, Geach2017}, mainly due to source
blending within the relatively large SCUBA beam and to insufficient corrections
for flux boosting \citep[Eddington bias;][]{Eddington1913}. Nevertheless,
sub-mm data are still challenging for semi-analytic galaxy formation models
(SAMs). In particular, SAMs generally under-predict the bright end of the
star formation rate (SFR) function at $z\simgt 2$ \citep{Niemi2012,
Gruppioni2015}.

A further challenge came from the first searches for sub-mm selected $z\simgt
4$ galaxies \citep{Dowell2014, Asboth2016, Ivison2016}: the inferred
surface densities at flux densities larger than several tens mJy at $500\,\mu$m
were found to be in excess of model predictions by about one order of
magnitude. More recent studies \citep{Donevski2018, Duivenvoorden2018} suggest
that the discrepancy with models may be accounted for by flux boosting due to
instrumental noise and confusion (including the contribution from clustering),
plus source blending. However, the issue is not settled yet (see
Section~\ref{sect:high_z_SFGs}).

On the other hand, a key ingredient to understand the star formation
process is the initial mass function (IMF), i.e., the relative number of newly
formed stars as a function of their mass.
It is still being debated whether or not the IMF is universal or varies with
conditions in star-forming regions. \citet{Bastian2010}, in their comprehensive
review, concluded that the data available at the time did not provide ``clear
evidence that the IMF varies strongly and systematically as a function of
initial conditions after the first few generations of stars''.

On the theoretical side, \citet{Padoan1997} presented a model whereby
the IMF depends on the average physical parameters (e.g., temperature, density,
and velocity dispersion) of the star formation sites\footnote{Note that
in \citet{Padoan1997}, ``universality of the IMF'' has a different meaning than
in this paper. Following \citet{Bastian2010}, by ``universal IMF'' we mean
``IMF independent of conditions in star formation sites''. Instead, the IMF by
\citet{Padoan1997} is universal in the sense that its functional form is a
consequence of the statistics of random supersonic flows but varies with local
properties. }. The model predicts a larger fraction of massive stars (i.e., a
top-heavier IMF) in starbursts. In starburst galaxies the cosmic ray energy
densities generated by the deaths of the most massive stars may raise the gas
temperatures even in the coolest star-forming clouds, thus increasing the Jeans
mass, hence the characteristic stellar mass, i.e., the IMF shape
\citep{Papadopoulos2011}. \citet{Chiosi1998} argued that this model
alleviates or solves some difficulties faced by models for the formation and
evolution of elliptical galaxies.

Recently, observational evidence of a top-heavy IMF (i.e., of an IMF
with more massive stars than expected from canonical IMFs, such as that by
\citealt{Chabrier2003} or by \citealt{Kroupa2013}) have been reported to
account for the low $^{13}$C/$^{18}$O abundance ratio found in both local
ultraluminous infrared galaxies \citep{Sliwa2017,BrownWilson2019} and in four
gravitational lensed sub-mm galaxies at $z\sim 2$--3 \citep{Zhang2018}.

A greater abundance of massive stars at high-$z$ translates into a higher
sub-mm luminosity of dusty galaxies at fixed SFR, i.e., into a greater
abundance of sub--mm bright galaxies than expected under the assumption of a
universal IMF. How do the counts of high-$z$ galaxies implied by the
top-heavier IMFs proposed by \citet{Zhang2018} compare with observations?

To address this question we need a physical model, linking the IR luminosity of
galaxies to their star formation history, taking also into account the effect
of active galactic nuclei (AGNs) growing in their centers. The physical model
by \citet{Cai2013}, briefly described in Section~\ref{sect:model}, allows us to
carry out this investigation. The baseline model adopts a \citet{Chabrier2003}
IMF, assumed to be universal. This IMF is described by a broken
power-law: $dN/d\log m \propto m^{-\alpha_i}$ with $\alpha_1 = 0.4$ for $0.1
\le m/M_\odot \le 1$ and $\alpha_2 = 1.35$ for $1 \le m/M_\odot \le 100$, where
$m$ is the stellar mass and $dN$ the number of stars within $d\log m$.

An extremely top-heavy IMF ($dN/d\log m = \hbox{constant}$) was
previously proposed by \citet{Baugh2005} to bring the counts of sub-mm
galaxies predicted by their semi-analytic model into agreement with
observations. Less top-heavy IMFs have been invoked by \citet{Lacey2016} and
\citet{Cowley2019} to predict sub-mm counts. However other models
\citep[e.g.,][]{Granato2004, Cai2013} accounted for the sub-mm counts adopting
a standard \citep{Chabrier2003} IMF. None of these models considered the counts
of sub-mm selected galaxies at $z\simgt 4$ or $z\simgt 6$.

In Section~\ref{sect:IMF} we investigate the effect of a top-heavier than
Chabrier IMF on several observables. In Section~\ref{sect:conclusions} we summarize
and discuss our main conclusions. Throughout this paper we adopt a flat $\Lambda$CDM
cosmology with the latest values of the parameters derived from Planck Cosmic
Microwave Background (CMB) power spectra: $H_0 = 67.4\,\hbox{km}\,\hbox{s}^{-1}\,
\hbox{Mpc}^{-1}$ and $\Omega_{\rm m} = 0.315$
\citep{PlanckCollaboration2018parameters}.

\section{Dusty star-forming galaxies at $z > 4$}\label{sect:high_z_SFGs}

\subsection{Summary of relevant observations}

The sub-mm extragalactic surveys with the \textit{Herschel} Space
Observatory at wavelengths extending over the peak of the dust emission
spectrum up to high redshifts, have made possible to select candidate high-$z$
galaxies. The first searches for $z\simgt 4$ galaxies were carried out by
\citet{Dowell2014}, \citet{Asboth2016}, and \citet{Ivison2016}. They used
\textit{Herschel} Spectral and Photometric Imaging Receiver (SPIRE) flux
densities at 250, 350, and $500\,\mu$m to search for sources with red colors,
indicative of high $z$. Since the spectral energy distribution (SED) of local
ultraluminous dusty star-forming galaxies typically peaks at rest-frame
wavelengths $\lambda \sim 100\,\mu$m, the SEDs of $z\simgt 4$ dusty
star-forming galaxies are expected to typically peak at $\lambda \simgt
500\,\mu$m in the observer's frame ($500\,\mu\rm m$ risers).

On this basis \citet{Dowell2014} selected sources having
$S_{500\,\mu\rm m} \ge S_{350\,\mu\rm m} \ge S_{250\,\mu\rm m}$ and
$S_{500\,\mu\rm m} \ge 30\,$mJy in three fields, part of the \textit{Herschel}
Multi-tiered Extragalactic Survey \citep[HerMES;][]{Oliver2012}, totaling an
area of $21\,\hbox{deg}^2$. Follow-up spectroscopy of the first 5 of the 38
sources selected in this way showed that 4 have $z$ in the range 4.3--6.3 and
the fifth has $z=3.4$, confirming their method is efficient at selecting
high-$z$ dusty galaxies. After correcting for completeness and purity, the
surface density of risers with $S_{500\,\mu\rm m} \ge 30\,$mJy was found to be
$3.3\pm 0.8\,\hbox{deg}^{-2}$. For comparison, the \citet{Cai2013} model
predicted $0.6\,\hbox{deg}^{-2}$ galaxies above $z=4$ at this flux density
limit. {The discrepancy with the \citet{Bethermin2017} model amounts to about a
factor of 10.}

\citet{Asboth2016} adopted the same search technique over a much larger
area ($274\,\hbox{deg}^2$ of the HerMES Large Mode Survey (HeLMS) field), with
a shallower flux density limit ($S_{500\,\mu\rm m} \ge 52\,$mJy). Measurements
of the spectroscopic redshifts of 2 of their 477 galaxies yielded $z=5.1$ and 3.8.
The size of their sample allowed them to compute, for the first time,
differential number counts. By means of Monte Carlo simulations they estimated
the intrinsic number counts, taking into account corrections for completeness,
Eddington bias, false detections, and blending effects. Their results are
consistent with the excess reported by \citet{Dowell2014}.

\citet{Ivison2016} selected sources detected over the $\simeq
600\,\hbox{deg}^2$ of the \textit{Herschel} Astrophysical Terahertz Large Area
Survey \citep[H-ATLAS;][]{Eales2010} with $S_{500\,\mu\rm m} \ge 30\,$mJy,
$S_{500\,\mu\rm m}/S_{250\,\mu\rm m} \ge 1.5$, and $S_{500\,\mu\rm
m}/S_{350\,\mu\rm m} \ge 0.85$. Sub-mm images with SCUBA-2 and with LABOCA were
obtained for a suitably chosen subset of these sources. Photometric redshifts
were estimated comparing the observed sub-mm SEDs with those of three well
sampled SED templates, found to be representatives of high-$z$ dusty
star-forming galaxies. In this way, the redshift distribution was derived. The
fraction of $z>4$ galaxies was found to be $32\pm 5\%$. Their surface density
is $2.1\pm 0.6\,\hbox{deg}^{-2}$, somewhat lower than, but consistent with the
estimate by \citet{Dowell2014} and substantially higher than the
\citet{Cai2013} and {the \citet{Bethermin2017} predictions}.

\begin{figure*}[tb!]
\centering
\includegraphics[width=0.49\textwidth]{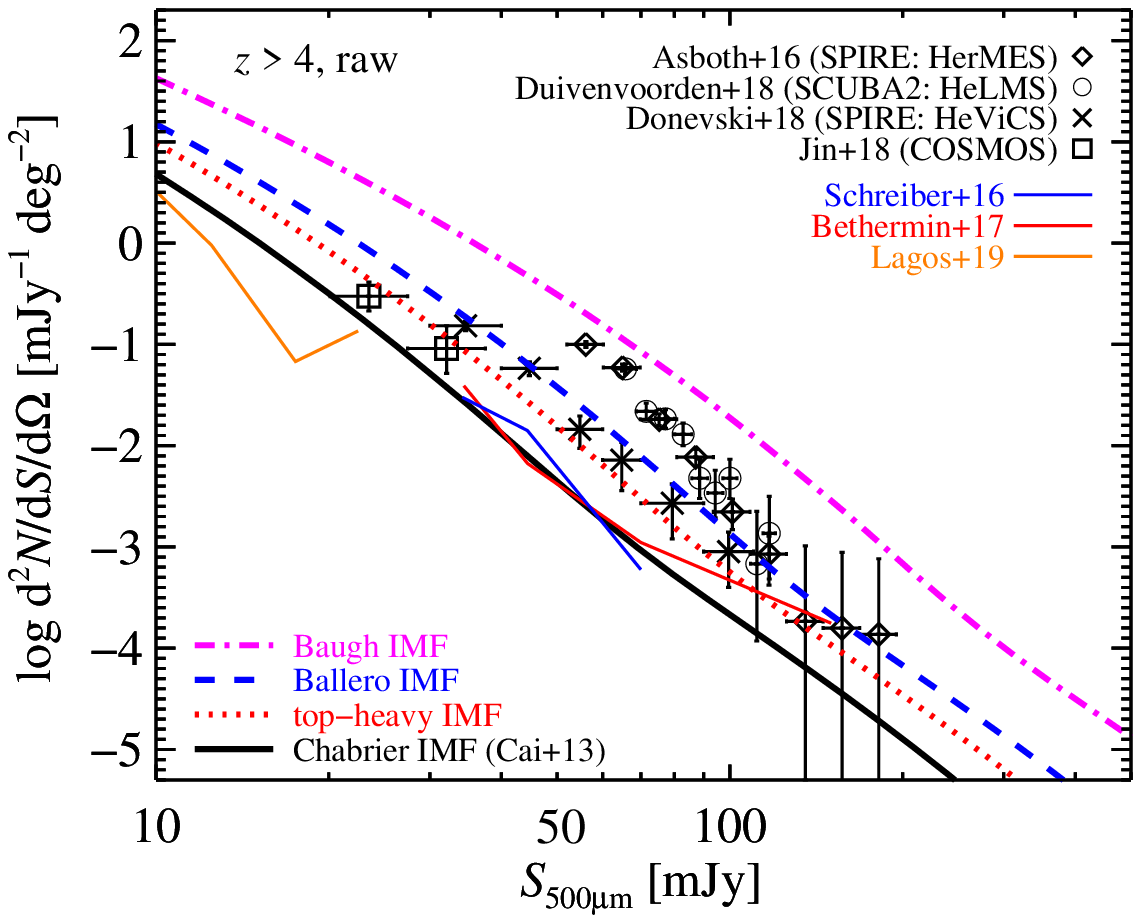}
\includegraphics[width=0.49\textwidth]{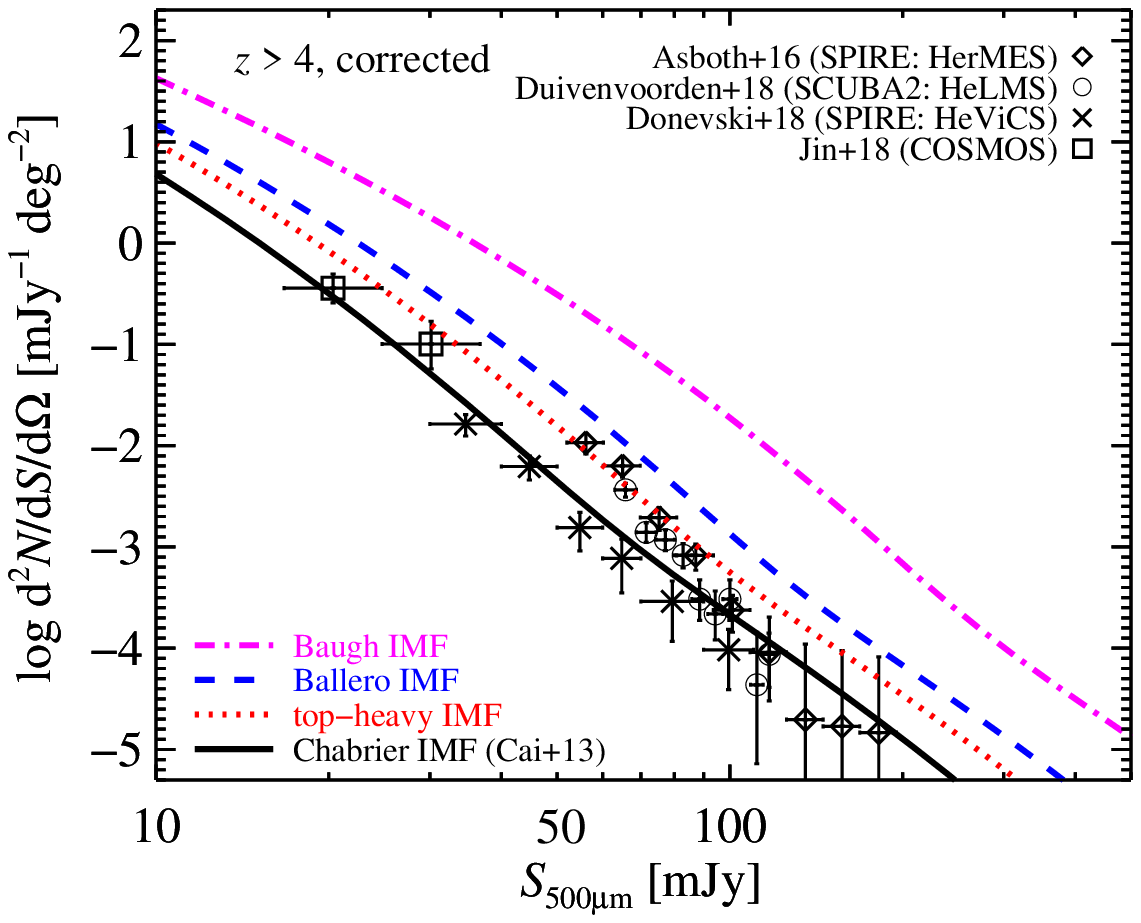}
\caption{
{Left panel:} raw differential $500\,\mu$m counts of $z\simgt 4$ galaxies estimated by
\citet{Asboth2016}, \citet{Duivenvoorden2018}, \citet{Donevski2018}, and \citet{Jin2018},
compared with model predictions. We have derived the \citet{Jin2018} counts extracting
from the published catalogue sources with $S_{\rm 500\mu\rm m}> 20\,$mJy and
signal to noise ratio $\ge 5$. \citet{Dowell2014} and \citet{Ivison2016}  reported only
integral counts. Therefore their results cannot be plotted here. They are consistent with \citet{Asboth2016}.
The thick solid black line shows the total counts yielded
by the baseline physical model by \citet{Cai2013} with updated parameters for the ``Chabrier'' IMF
(see Section~\ref{sect:IMF}).
Counts for other choices of the IMF, including the
``top-heavy'' (dotted red line), ``Ballero'' (dashed blue line), and ``Baugh''
(dot-dashed magenta line), are also shown (see Section~\ref{sect:IMF}).
In addition, we show the predictions of the phenomenological models by \citet[][thin solid blue line]{Schreiber2016} and
\citet[][thin solid red line]{Bethermin2017},
taken from the left panel of Figure~9 of \citet{Donevski2018}, as well as the prediction of 
\citet[][thin solid orange line; private communication]{Lagos2019} using the SHARK SAM \citep{Lagos2018}.
{Right panel:} Differential counts of $z\simgt 4$ galaxies corrected for
the effect of noise and source blending. For the counts by \citet{Asboth2016}
and \citet{Donevski2018} we have adopted the correction factor of 168/18
estimated by \citet[][cf. their Section~6.2]{Bethermin2017}, assuming a factor
of 2 uncertainty in the correction factor. \citet[][cf. their
Section~4.1.2]{Duivenvoorden2018} have provided their own estimates for the
correction factor [$(172 \pm 18)/11$]. Since \citet{Jin2018} have used
``super-deblended'' photometry, we have only applied to each source the
correction for flux boosting given by Equation~(\protect\ref{eq:HT}).
The lines are the same as in the left panel.
}\label{fig:zgt4diffcounts}
\end{figure*}

{However, as suggested by more recent studies \citep{Donevski2018, Duivenvoorden2018}, 
the discrepancy with models may be accounted for by flux boosting and source blending.}
\citet{Donevski2018} used the criteria of \citet{Dowell2014} to select
$500\,\mu\rm m$ risers over $55\,\hbox{deg}^2$ of the \textit{Herschel} Virgo
Cluster Survey \citep[HeViCS;][]{Davies2010} down to a deeper flux density
limit at $250\,\mu$m: $S_{250\,\mu\rm m} \ge 13.2\,$mJy. They extracted sources
from multi-wavelength maps with a code combining priors on positions and SEDs.
\citet{Duivenvoorden2018} performed SCUBA-2 follow-up observations of 188 of
the brightest 200 sources ($S_{500\,\mu\rm m} \ge 63\,$mJy) selected by
\citet{Asboth2016} and estimated their redshifts by SED fitting. Both groups
used model-based simulations to correct the raw counts for the effects of flux
boosting and of source blending. On this footing they found consistency with
model predictions (see Figure~\ref{fig:zgt4diffcounts}).

\citet{Jin2018} have published a ``super-deblended'' far-infrared to
(sub)mm photometric catalogue in the Cosmic Evolution Survey (COSMOS) field
with spectroscopic or photometric redshifts. We have extracted from their
catalogue $z> 4$ sources with $S_{\rm 500\mu\rm m}> 20\,$mJy and signal to
noise ratio $\ge 5$ and derived their differential number counts after applying
the correction for flux boosting (cf. Equation~(\protect\ref{eq:HT})). Again, the
corrected counts are consistent with model predictions.

\subsection{Potential observational biases}

A strong effect of flux boosting on the counts of $z\simgt 4$ galaxies is
indeed expected as a consequence of their steep slope. As shown by
\citet{HoggTurner1998}, in the case of Gaussian noise, the ratio of the maximum
likelihood true flux, $S_{\rm ML}$, to the observed flux, $S_{\rm o}$,  is
related to the slope of the integral counts, $q$, and to the signal-to-noise
ratio, $r$, by
\begin{equation}\label{eq:HT}
\frac{S_{\rm ML}}{S_{\rm o}} =\frac{1}{2}+\frac{1}{2}\left(1-\frac{4q+4}{r^2}\right)^{1/2}.
\end{equation}
The observationally determined counts shown in
Figure~\ref{fig:zgt4diffcounts} consistently indicate, for $z>4$ galaxies, $q\sim
3.6$, implying that there is no finite maximum likelihood value for $r<4.3$.
For $r=4.5$, close to the critical value $r=4.3$, and $r=5$, usually
taken as the threshold for reliable detections, we have $S_{\rm ML}/S_{\rm
o}=0.651$ and 0.757, respectively. This bias in the measured flux density
translates in a bias in the integral source counts of $(S_{\rm ML}/S_{\rm
o})^{-q}$, i.e., of factors of 4.7 and 2.7, respectively, for the two values of
$r$. Note that these are just examples to illustrate that noise has
indeed a strong impact on observed counts.

This simple estimate does not account for the non-Gaussian, spatially
correlated fluctuations due to clustering as well as for the complexity of
source selection. 
Both \citet{Donevski2018} and \citet{Duivenvoorden2018} have extensively used
simulations from \citet{Bethermin2017} to evaluate their selection method.
To deal with these effects, \citet{Bethermin2017}
have produced a $2\,\hbox{deg}^2$ simulation of the extragalactic sky from
$24\,\mu$m to 1.3\,mm, called SIDES (Simulated Infrared Dusty Extragalactic
Sky). The simulation uses an updated version of the empirical model by
\citet{Bethermin2012,Bethermin2015} and includes clustering based on the spatial distribution
of dark matter halos yielded by a state-of-the-art $N$-body simulation; halos are
populated with galaxies using abundance matching. Since a $2\,\hbox{deg}^2$
cone is too small to provide enough statistics on the rare $z>4$ sources
selected by most of the studies mentioned above, \citet{Bethermin2017} built
also a $274\,\hbox{deg}^2$ simulation, which however does not contain
clustering. The simulation of the full source extraction showed that the
combination of noise, source blending, and weak lensing can dramatically boost
the number density of red sources selected with the criteria by
\citet{Asboth2016}, compared to their real number density. This conclusion
contradicts that by \citet{Asboth2016} who also used simulations to support
their surface density estimates. \citet{Bethermin2017} explained the
contradiction claiming that they have identified a potentially incorrect
assumption in the  simulation by \citet{Asboth2016}.

\citet{Bethermin2017} attributed a substantial contribution to the
number density overestimate to an overestimate of source flux densities at
$500\,\mu$m due to multiple sources within the relatively large
\textit{Herschel} beam (source blending) at this wavelength; according to their
simulation only $\sim 60\%$ of the measured flux density comes from the
brightest source within the beam. Accounting for these effects can reconcile
observations with model predictions. \citet{Bethermin2017} however acknowledged
that correcting the observed counts for all the observational effects is a
complex task and that their approach has limitations.

The issue is not settled, however. None of the galaxies, drawn from the sample
of $z\simgt 4$ candidates selected by \citet{Ivison2016}, observed by
\citet{Fudamoto2017} in 1.3-- or 3--mm continuum using the Northern
Extended Millimeter Array (NOEMA) and Atacama Large Millimeter Array (ALMA)
were found to be affected by blending, as would be expected from
simulations of \citet{Bethermin2017}; multiplicity may affect at most the $\sim
20$ per cent of  their targets that remained undetected in continuum.

\begin{figure}[tb!]
\centering
\includegraphics[width=0.45\textwidth]{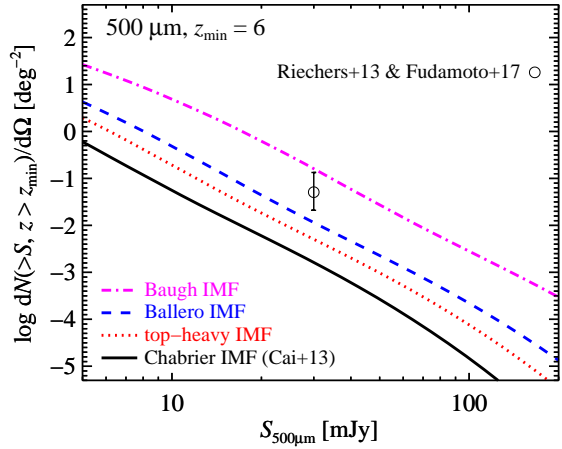}
\caption{Predicted integral $500\,\mu\rm m$ counts of dusty galaxies at $z>6$
for different choices of the IMFs, namely the ``Chabrier'' (baseline model; solid black line),
the ``top-heavy'' (dotted red line), the ``Ballero'' (dashed blue line),
and the ``Baugh'' (dot-dashed magenta line) IMF (see Section~\ref{sect:IMF}). The data point shows
the average surface density for $S_{500\mu\rm m}>30\,$mJy derived from searches
by \citet{Riechers2013} and \citet{Fudamoto2017}.
}\label{fig:zgt6intcounts}
\end{figure}

Further hints of a high-$z$ excess come from searches of $z>6$ dusty galaxies.
\citet{Riechers2013} found a galaxy at $z=6.34$ in a sample with ultra-red
\textit{Herschel}/SPIRE colours with $S_{500\mu\rm m}>30\,$mJy over an area of
$21\,\hbox{deg}^2$. The corresponding surface density is $N(>S_{500\mu\rm m}>
30\,\hbox{mJy}, z>6)=0.048^{+0.110}_{-0.039}\,\hbox{deg}^{-2}$ (Poisson
errors). A second $z>6$ galaxy (at $z =6.027$) was reported by
\citet{Fudamoto2017} and \citet{Zavala2018}. This galaxy belongs to a sample of the reddest
\textit{Herschel}/SPIRE galaxies over the H-ATLAS area ($660\,\hbox{deg}^2$),
again with $S_{500\mu\rm m}>30\,$mJy. Taking into account the correction for
the incompleteness of the parent sample \citep[a factor of
$1/0.028$;][]{Ivison2016} we get $N(>S_{500\mu\rm m}> 30\,\hbox{mJy},
z>6)=0.054^{+0.125}_{-0.045}\,\hbox{deg}^{-2}$. The average of the two
estimates is $N(>S_{500\mu\rm m}> 30\,\hbox{mJy},
z>6)=0.051^{+0.083}_{-0.030}\,\hbox{deg}^{-2}$ (see
Figure~\ref{fig:zgt6intcounts}).

%%%%%%%%%%%%%%%%%%%%%%%%%%%%%%%%%%%%%%%%%%%%%%%%%%%%%%%%%%%%%%%%%%%%%%%%%%%%%%%%
\section{Model}\label{sect:model}

{While at $z\simlt 1.5$ star formation mostly occurs in late-type
galaxies (spiral, irregular, and disky starburst galaxies), at} the high
redshifts of interest here the {observed IR/sub-mm} galaxy luminosity
functions are dominated by proto-spheroids, i.e., by ellipticals and bulges of
disk galaxies in the process of forming the bulk of their stars
{\citep[e.g.,][]{Granato2004, Thomas2010, SilkMamon2012}}. The
\citet{Cai2013} model provides a physically grounded description of their
formation and evolution, building on the work by \citet{Granato2004},
\citet{Lapi2006}, and \citet{Lapi2011}.

{The model adopts the halo formation rate, $dN(M_{\rm H},z)/dt$, as a
function of redshift, $z$, and of halo mass, $M_{\rm H}$, yielded by
large-scale $N$-body simulations. For $z\simgt 1.5$ the $dN(M_{\rm H},z)/dt$ is
well approximated by the positive term of the derivative with respect to cosmic
time of the epoch-dependent halo mass function. For the latter, the analytical
approximation by \citet{ShethTormen1999} was used. High resolution $N$-body
simulations \citep[e.g.,][]{Wang2011} have identified two phases of the halo
evolution. A first fast collapse phase, including major mergers, forms the bulk
of its mass. The following phase produces a slow growth of the halo outskirts
by many minor mergers and diffuse accretion. This second stage has little
effect on the inner potential well where the visible galaxy resides. Based on
these results, the model assumes that the star formation and the AGN growth are
basically regulated by in-situ processes, as confirmed by subsequent
high-resolution simulations \citep[e.g.,][]{Pillepich2015}.} The star formation
history of proto-spheroids {with any halo mass and formation redshift}
and the accretion history onto the central super-massive black holes are
computed by solving a set of equations describing the evolution of gas phases
and of the active nucleus, including the effect of cooling, condensation into
stars, radiation drag, accretion, and feedback from supernovae (SNe) and from
the AGN.

These equations yield the SFR of each galaxy and the bolometric luminosity of
the AGN as a function of halo mass, formation redshift, and galactic age.
{The SFR was converted to the total infrared (IR; 8--$1000\,\mu$m) luminosity using
the standard calibration \citep[e.g.,][]{KennicuttEvans2012}.} The
epoch-dependent {bolometric} luminosity functions of galaxies, of AGNs,
and of objects as a whole (galaxy plus AGN) are computed coupling luminosities
with the halo formation rate.

{To derive monochromatic luminosity functions, the SED of the strongly
lensed $z\simeq 2.3$ galaxy SMM~J2135-0102 \citep{Swinbank2010}, modeled using
the GRASIL software package \citep{Silva1998} was adopted for the dusty galaxy
component. This SED was found to be representative of high-$z$ dusty galaxies
\citep{GonzalezNuevo2012, Ivison2016}. For AGNs two main phases of the
co-evolution with the galaxy component were considered. For the first phase,
when the black-hole growth is enshrouded by the dusty interstellar medium (ISM)
of the host galaxy, a heavily absorbed AGN SED, taken from the library by
\citet{GranatoDanese1994}, was adopted. For the second phase, when the AGN
shines after having swept out the galaxy ISM, we adopted the mean QSO SED by
\citet{Richards2006}, extended to sub-mm wavelengths assuming a grey-body
emission with dust temperature $T_{\rm dust} = 80\, \rm K$ and emissivity index
$\beta = 1.8$.}

{After determining the values of the parameters from a comparison with a
set of multi-frequency luminosity functions \citep[see][for details]{Cai2013}
the model successfully reproduced} a broad variety of data: multi-frequency and
multi-epoch luminosity functions of galaxies and AGNs {beyond those used
for the model calibration,} counts of unlensed and gravitationally lensed
sources, redshift distributions, and clustering properties \citep{Xia2012,
Cai2013, Cai2014, Carniani2015, Bonato2017}. In particular the model accounts
for the redshift-dependent IR luminosity function
determined by \citet{Gruppioni2013} up to $z\simeq 4$ \citep{Bonato2014a} and
fits the most recent mm and sub-mm counts and redshift distributions
\citep{Bonato2019, Gralla2019, DeZotti2019} as well as the SED and the power
spectrum of the cosmic infrared background \citep[][]{Cai2013}.

\section{A top-heavier IMF?}\label{sect:IMF}

The bolometric luminosity {of a young stellar population is dominated by
massive, short-lived, UV-producing stars \citep[e.g.,][]{Kennicutt1998,
Madau1998}. Hence it} is approximately proportional to the mass fraction in
massive ($m \ge 8\,M_\odot$) stars. For the heavily dust obscured galaxies of
interest here, $L_{\rm IR}\simeq L_{\rm bol}$. The calibration of the
relationship between $L_{\rm IR}$ and SFR for complete dust obscuration adopted
by \citet{Cai2013} refers to the Chabrier IMF {\citep{Chabrier2003}
whose power-law indices below and above $1\,M_\odot$ are -0.4 and -1.35,
respectively (in terms of $dN/d\log m$)}.
It writes
\begin{equation}\label{eq:calib}
\frac{L_{\rm IR}}{\hbox{erg}\,\hbox{s}^{-1}}=k_{\star,\rm IR}\times 10^{43}
\frac{\hbox{SFR}}{M_\odot\,\hbox{yr}^{-1}}
\end{equation}
with $k_{\star,\rm IR}=3.1$. The IMFs suggested by \citet{Zhang2018} have
{power-law indices below and above $0.5\,M_\odot$, respectively, of -0.3
and -1.1 for the ``top-heavy'' IMF \citep{Zhang2018} and of -0.3 and -0.95 for
the ``Ballero'' IMF \citep[][]{Ballero2007}. }

The mass fractions in the 8--$100\,M_\odot$ range are of 23\%, 33\%, and 44\%
for the ``Chabrier'', ``top-heavy'', and ``Ballero'' IMF, respectively. Thus
the bolometric luminosity at fixed SFR increases, compared to the ``Chabrier''
case, by factors of $\simeq 1.4$ and of $\simeq 1.9$ for the ``top-heavy'' and
``Ballero'' IMF, respectively. In the extreme case of the flat IMF
{(i.e., $dN/d\log m = \hbox{constant}$)} used by \citet{Baugh2005} for
starbursts, the mass fraction in $\ge 8\,M_\odot$ stars increases by a factor
$\simeq 4$.

\citet{Zhang2018} mention a factor of 6--7 increase in the mass fraction of
massive stars moving from the ``Kroupa'' \citep{Kroupa2013} to the ``Ballero''
IMF. This may look in conflict with the substantially lower factor we
find for the ``Chabrier'' case since the ``Kroupa'' IMF is generally considered
to give a relation between $L_{\rm IR}$ and SFR almost identical to the
``Chabrier'' IMF \citep[e.g.,][]{ChomiukPovich2011}. {However}, \citet{Zhang2018}
refer to a version of the Kroupa IMF with a much lower fraction of massive
stars {(8.6\%)}\footnote{The fraction of massive stars ($8-100~M_\sun$)  reported in Table~1 of \citet{Zhang2018}  for the ``Kroupa'' IMF (6.9\%) is a typo which does not affect their chemical evolution models  (Zhi-Yu Zhang, private communication). With the corrected fraction of massive stars, 8.6\%, the increase from their ``Kroupa'' to the ``Ballero'' IMF is a factor of $\simeq 5$.} than that (23\%) yielded by the IMF
adopted by \citet{Cai2013}.

The IMFs impinges also on other quantities: the supernova (SN) rate and the
fraction, ${\cal R}$, of the mass of a stellar generation that returns to the
interstellar medium (restitution fraction). Both factors affect the evolution
of the metallicity and of the SFR. An increase of ${\cal R}$ makes more
material available for star formation while feedback effects are obviously
stronger for higher SN rates, with the effect of decreasing the SFR.

SNe dominate feedback effects not only for less massive galaxies but also for
most of the intense star formation phase of massive galaxies. The AGN feedback
takes over in the latter objects only during the last few e-folding times of
the central black hole growth. The fraction of stars per stellar generation
that explode as SNe, i.e., the ratio of the number of stars more massive than
$8\,M_\odot$ to the total number of stars, is $\simeq 1\%$ for the ``Chabrier''
IMF and increases to $\simeq 1.4\%$, 2.2\%, and 37\% for the ``top-heavy'',
``Ballero'', and ``Baugh'' IMFs, respectively. The corresponding parameter
defined in \citet{Cai2013} is the number of SNe per unit solar mass of
condensed stars: $N_{\rm SN} \simeq 1.2 \times 10^{-2}/M_\sun$, $1.6 \times
10^{-2}/M_\sun$, $2.0 \times 10^{-2}/M_\sun$, and $2.5 \times 10^{-2}/M_\sun$,
for the ``Chabrier'', ``top-heavy'', ``Ballero'', and ``Baugh'' IMFs,
respectively.

The restitution fraction, ${\cal R}$, affects the SFR and the chemical
evolution. It depends on the minimum mass, $M_{\rm min}$, of stars with
lifetimes not exceeding the age of the universe at the considered redshift. We
set $M_{\rm min}=2\,M_\odot$ since the lifetime of a star of that mass with
solar metallicity is slightly lower than the age of the universe at $z=4$
($\simeq 1.5\,$Gyr) for our choice of cosmological parameters. For $M_{\rm
min}=2\,M_\odot$, ${\cal R}=0.393$, 0.466, 0.56, and 0.83 for the ``Chabrier'',
``top-heavy'', ``Ballero'', and ``Baugh'' IMFs, respectively. {After
having checked that the results are very weakly sensitive to the choice for
$M_{\rm min}$, we have neglected its dependence on metallicity and used the
same value for proto-spheroids at all redshifts. Note however that at $z\simlt
1.5$ the dominant star-forming population are late-type galaxies, for which we
have kept the same model used by \citet{Cai2013}. }

For each of the considered IMFs we have re-run the code solving the set of
equations written down by \citet{Cai2013}. To compute observable quantities
such as number counts or luminosity functions, the output of these
calculations, i.e., the SFR and the bolometric luminosity of the central AGN as
a function of halo mass, formation redshift, and galactic age, need to be
coupled with the SEDs of source populations.

\begin{figure}[tb!]
\centering
\includegraphics[width=0.5\textwidth]{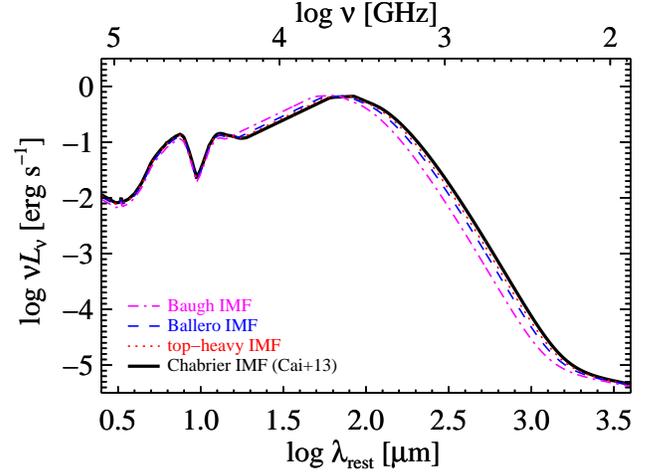}
\caption{
Galaxy SEDs, with total IR luminosity normalized to unit, associated to
the ``Chabrier'', ``top-heavy'', ``Ballero'', and ``Baugh'' IMFs of proto-spheroidal
galaxies.}
\label{fig:SED}
\end{figure}

We have kept the AGN SEDs used by \citet{Cai2013}. In the case of dust
luminosities powered by star formation, we need to take into account that
higher luminosities entail, {on average}, higher dust temperatures,
i.e., warmer SEDs {\citep{Schreiber2018, Liang2019}}. Since a
sophisticated treatment is not warranted in this exploratory study we have
taken this into account by assuming that the effective dust temperature,
$T_{\rm d}$, hence the peak frequency of the dust emission, $\nu_{\rm peak}$,
scales as $L_{\rm IR}^{1/6}$ \citep[e.g.,][]{Liang2019}. Then, for different
IMFs, $\nu_{\rm peak} \propto k_{\star,\rm IR}^{1/6}$; the full continuum SED
was shifted in frequency by the corresponding factor. Figure~\ref{fig:SED}
shows the SEDs associated to the considered IMFs. {This approach
assumes, for each choice of the IMF, the same \textit{effective} SED for all
galaxies. This is clearly an approximation, necessary to keep the model
manageable and justified by its capability of accounting for a broad variety of
data.  }

\citet{Cai2013} adopted $N_{\rm SN} \simeq 1.4 \times 10^{-2}/M_\sun$ and
${\cal R}=0.54$. {These values differ from the values given above for
the ``Chabrier'' IMF because of a slightly different way of computing $N_{\rm
SN}$ and of the choice of a lower $M_{\rm min}$ ($1\,M_\odot$ instead of
$2\,M_\odot$) motivated by the fact that the focus was on lower redshifts.} We
have re-run the model with {the new} values and used the results of the
updated model. The differences in the derived IR luminosity functions for the
``Chabrier'' IMF turned out to be irrelevant.

{Figure~\ref{fig:zgt4diffcounts} shows that the observed \textit{raw}
$500\,\mu$m counts of $z\simgt 4$ galaxies are consistently and substantially
above the predictions of the baseline \citet{Cai2013} model, which adopted a
universal \citet{Chabrier2003} IMF. Taken at face value they strongly favor a
top-heavier IMF such as the ``top-heavy'' or the ``Ballero''; the extreme
``Baugh'' IMF is however inconsistent with the data. A similar indication comes
from the estimate at $z> 6$ (Figure~\ref{fig:zgt6intcounts}). The correction
factors derived from the \citet{Bethermin2017} simulations bring the $z\simgt
4$ data close to agreement with the baseline model. However, several of the
corrected data points are still consistent, within the uncertainties, with a
top-heavy IMF; the main exception to that are the counts by
\citet{Donevski2018}. High resolution follow-up of the two $z>6$ galaxies did
not show indications of blending, so that no substantial correction appears to
be necessary. However the statistical uncertainty of this data point is large.}

\begin{figure*}%[tb!]
\centering
\includegraphics[width=\textwidth]{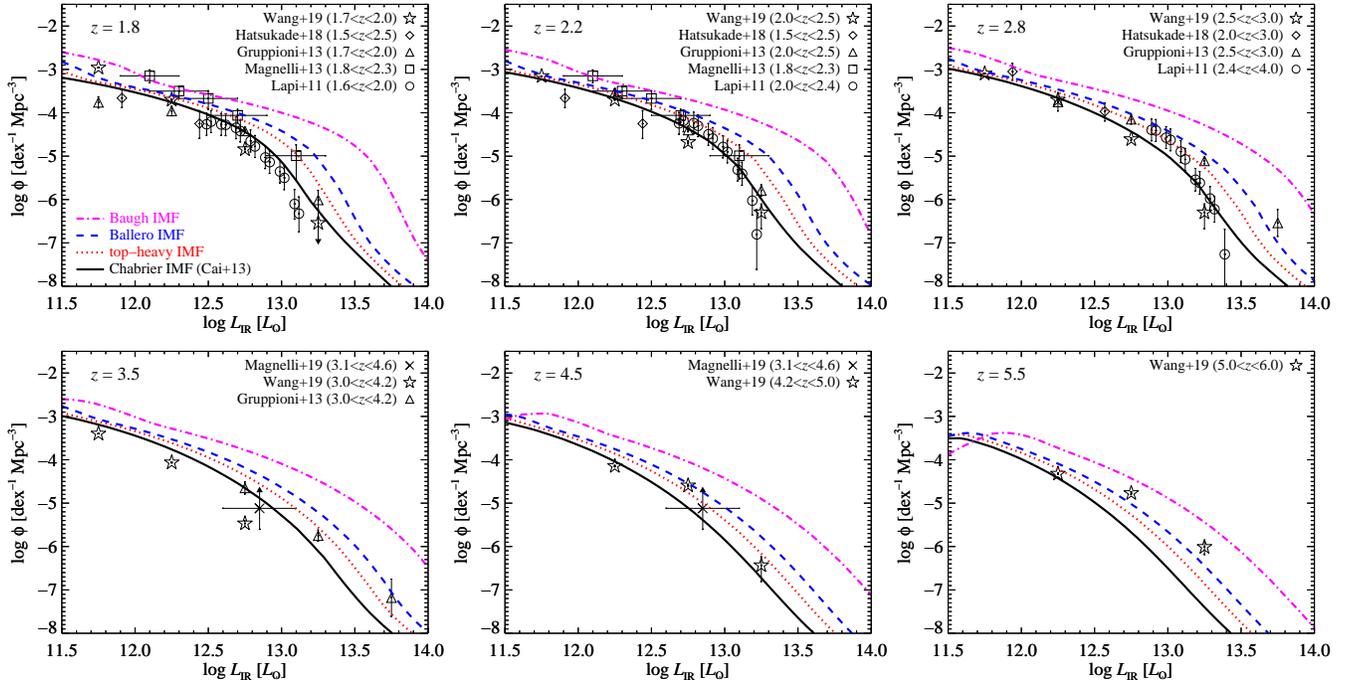}
\caption{Comparison of the most recent
determinations of the high-$z$ IR luminosity functions of galaxies with the
model by \citet{Cai2013} for different choices for the IMF of high-$z$ proto-spheroidal galaxies:
a universal ``Chabrier'' IMF (baseline model; $k_{\star,\rm IR}=3.1$; {$k_{\star,\rm IR}$
is defined by Equation~(\ref{eq:calib})});
the ``top-heavy'' ($k_{\star,\rm IR}=3.1 \times 1.4$) and the ``Ballero''
($k_{\star,\rm IR}=3.1 \times 1.9$) IMFs suggested by \citet{Zhang2018};
the extreme, flat, IMF ($k_{\star,\rm IR}=3.1 \times 4$) used by \citet{Baugh2005}.
}\label{fig:LF_IR}
\end{figure*}

{Figure~\ref{fig:LF_IR} compares the observed total IR ($8-1000\,\mu$m)
luminosity functions at $z=1.8$ to $z=5.5$ with results for the baseline
\citet{Cai2013} model (black solid lines) and the other choices for the IMF.
The comparisons with observational estimates of \citet{Magnelli2013},
\citet{Hatsukade2018}, \citet{Magnelli2019}, and \citet{Wang2019} are new. The
samples used by \citet{Gruppioni2013} and by \citet{Magnelli2013} were drawn
from deep surveys with the \textit{Herschel} Photodetector Array Camera and
Spectrometer (PACS) which has a better angular resolution than
\textit{Herschel}/SPIRE. These data are therefore much less affected by
confusion and blending problems.}

{\citet{Wang2019} used \textit{Herschel}/SPIRE data in the COSMOS field
but exploited the rich, deep multi-frequency data available in this field and
the SIDES empirical model to deblend and deconfuse the \textit{Herschel}/SPIRE
images. In this way they were able to extend the sub-mm source counts by more
than a factor of 10 below the confusion limit and to extend estimates of the IR
luminosity function to fainter luminosities and out to higher redshifts (up to
$z\sim 6$) than \citet{Gruppioni2013}. The IR luminosity functions computed by
\citet{Hatsukade2018} and by \citet{Magnelli2019} are based on high resolution
ALMA surveys and are therefore unaffected by confusion and blending.}

{The baseline model by \citet{Cai2013} yields results in remarkable
agreement with the luminosity function data, especially at $z\simeq 1.8$ and
$\simeq 2.2$. An excess shows up especially at the highest redshifts ($z\simeq
3.5$, 4.5, and 5.5). The significance of the discrepancy is hard to assess,
however, because the published errors are purely Poisson while much larger
uncertainties may be associated to the source selection, to photometric
redshifts, and to the deblended photometry. Errors preferentially shift sources
from more populated to less populated redshift/luminosity bins (an effect
analogous to the Eddington bias), implying that the number of rare highest
luminosity/redshift sources may be easily overestimated. However, taken at face
value, these data support the view of an excess of high-$z$ sub-mm galaxies
that may require a top-heavier IMF, such as the ``top-heavy'' or the
``Ballero'', to be accounted for. Again, the ``Baugh'' IMF implies source
densities well in excess of observations.}

\section{Conclusions}\label{sect:conclusions}

We have exploited the physical model by \citet{Cai2013} to investigate the
impact of a top-heavy IMF on sub-mm counts of high-$z$ ($z>4$ and $z>6$) dusty
galaxies.

The first estimates of sub-mm counts at $z\simgt 4$ galaxies \citep{Dowell2014,
Asboth2016, Ivison2016}, taken at face value, favor the top-heavier IMFs
suggested by \citet{Zhang2018}. On the contrary, the extreme (flat) IMF used by
\citet{Baugh2005} overpredicts, in the framework of the \citet{Cai2013}
approach, counts for $z>4$.

If the ``raw'' counts are corrected for flux boosting due to instrumental noise
and confusion following some recent studies \citep{Bethermin2017, Donevski2018,
Duivenvoorden2018}, {approximate} consistency with predictions based on
a universal ``Chabrier'' IMF is recovered. {On the other hand, with the
exception of the counts by \citet{Donevski2018}, the corrected data points are
also consistent, within the uncertainties, with a top-heavy IMF. A definite
conclusion on the abundance of $z>4$ sub-mm galaxies must await for higher
resolution data and a much higher fraction of spectroscopic redshifts.}

From the theoretical point of view an increase of the Jeans mass is expected in
the case of high SFRs due to the large cosmic ray energy densities which raise
the gas temperatures \citep{Papadopoulos2011}. A further increase of gas
temperatures with increasing $z$ is expected by effect of the raising CMB temperature \citep{Zhang2016}. The data indeed give
hints of an excess of $z>5$ sub-mm galaxies over expectations for a universal
IMF (Figure~\ref{fig:zgt6intcounts} and bottom-right panel of
Figure~\ref{fig:LF_IR}). {However they are liable to large errors,
arising from photometric redshift estimates, particularly tricky at these
redshifts, from model extrapolations to infer $L_{\rm IR}$ from mid-IR
photometry, from the de-blending process, and from field-to-field variations.
These errors are difficult to quantify and are not discussed by
\citet{Wang2019}. }

As suggested by \citet{Zhang2018}, top-heavier IMFs than usually assumed have
implications on SFRs derived from classical tracers such as $L_{\rm IR}$ or the
radio continuum luminosity. In turn, this {may} impact on the star formation history
of galaxies. {However we have found that top-heavier IMFs than Chabrier consistent with 
the high-$z$ data imply only a moderate increase of the $L_{\rm IR}/\hbox{SFR}$ 
ratio (by factors of 1.4—-1.9). 
Nevertheless, it is clearly important to assess the evolution, if any, of the IMF with cosmic time.}

In this paper we shown that the data already imply interesting constraints on
the IMF of high-$z$ dusty galaxies. Forthcoming or planned sub-mm spectroscopic
surveys, such as those with the Cerro Chajnantor Atacama Telescope
\citep[CCAT-prime;][]{Stacey2018} or with the Origins Space Telescope
\citep[OST;][]{Leisawitz2018, Bonato2019}, will shed light on this issue.

\section*{Acknowledgements}

{We are grateful to the referee for a careful reading of the manuscript
and detailed comments,} to A. Bressan for very useful suggestions, {to Zhi-Yu Zhang and D. Romano for their comments, and
to S. Duivenvoorden for clarifications on their counts of $z>4$ galaxies}. ZYC
is supported by the National Science Foundation of China (grant Nos. 11890693,
11873045, and 11421303). GdZ acknowledges financial support by ASI/INAF
Agreement 2014-024-R.0 for the {\it Planck} LFI activity of Phase E2.

\bibliographystyle{aasjournal}
\bibliography{highz_submm_rev.bbl}

\begin{thebibliography}{}
\expandafter\ifx\csname natexlab\endcsname\relax\def\natexlab#1{#1}\fi
\providecommand{\url}[1]{\href{#1}{#1}}
\providecommand{\dodoi}[1]{doi:~\href{http://doi.org/#1}{\nolinkurl{#1}}}
\providecommand{\doeprint}[1]{\href{http://ascl.net/#1}{\nolinkurl{http://ascl.net/#1}}}
\providecommand{\doarXiv}[1]{\href{https://arxiv.org/abs/#1}{\nolinkurl{https://arxiv.org/abs/#1}}}

\bibitem[{{Asboth} {et~al.}(2016){Asboth}, {Conley}, {Sayers}, {B{\'e}thermin},
  {Chapman}, {Clements}, {Cooray}, {Dannerbauer}, {Farrah}, \&
  {Glenn}}]{Asboth2016}
{Asboth}, V., {Conley}, A., {Sayers}, J., {et~al.} 2016, \mnras, 462, 1989,
  \dodoi{10.1093/mnras/stw1769}

\bibitem[{{Ballero} {et~al.}(2007){Ballero}, {Kroupa}, \&
  {Matteucci}}]{Ballero2007}
{Ballero}, S.~K., {Kroupa}, P., \& {Matteucci}, F. 2007, \aap, 467, 117,
  \dodoi{10.1051/0004-6361:20066786}

\bibitem[{{Barger} {et~al.}(1998){Barger}, {Cowie}, {Sanders}, {Fulton},
  {Taniguchi}, {Sato}, {Kawara}, \& {Okuda}}]{Barger1998}
{Barger}, A.~J., {Cowie}, L.~L., {Sanders}, D.~B., {et~al.} 1998, \nat, 394,
  248, \dodoi{10.1038/28338}

\bibitem[{{Bastian} {et~al.}(2010){Bastian}, {Covey}, \& {Meyer}}]{Bastian2010}
{Bastian}, N., {Covey}, K.~R., \& {Meyer}, M.~R. 2010, \araa, 48, 339,
  \dodoi{10.1146/annurev-astro-082708-101642}

\bibitem[{{Baugh} {et~al.}(2005){Baugh}, {Lacey}, {Frenk}, {Granato}, {Silva},
  {Bressan}, {Benson}, \& {Cole}}]{Baugh2005}
{Baugh}, C.~M., {Lacey}, C.~G., {Frenk}, C.~S., {et~al.} 2005, \mnras, 356,
  1191, \dodoi{10.1111/j.1365-2966.2004.08553.x}

\bibitem[{{B{\'e}thermin} {et~al.}(2015){B{\'e}thermin}, {De Breuck},
  {Sargent}, \& {Daddi}}]{Bethermin2015}
{B{\'e}thermin}, M., {De Breuck}, C., {Sargent}, M., \& {Daddi}, E. 2015, \aap,
  576, L9, \dodoi{10.1051/0004-6361/201525718}

\bibitem[{{B{\'e}thermin} {et~al.}(2012){B{\'e}thermin}, {Daddi}, {Magdis},
  {Sargent}, {Hezaveh}, {Elbaz}, {Le Borgne}, {Mullaney}, {Pannella}, {Buat},
  {Charmandaris}, {Lagache}, \& {Scott}}]{Bethermin2012}
{B{\'e}thermin}, M., {Daddi}, E., {Magdis}, G., {et~al.} 2012, \apjl, 757, L23,
  \dodoi{10.1088/2041-8205/757/2/L23}

\bibitem[{{B{\'e}thermin} {et~al.}(2017){B{\'e}thermin}, {Wu}, {Lagache},
  {Davidzon}, {Ponthieu}, {Cousin}, {Wang}, {Dor{\'e}}, {Daddi}, \&
  {Lapi}}]{Bethermin2017}
{B{\'e}thermin}, M., {Wu}, H.-Y., {Lagache}, G., {et~al.} 2017, \aap, 607, A89,
  \dodoi{10.1051/0004-6361/201730866}

\bibitem[{{Bonato} {et~al.}(2014){Bonato}, {Negrello}, {Cai}, {De Zotti},
  {Bressan}, {Lapi}, {Gruppioni}, {Spinoglio}, \& {Danese}}]{Bonato2014a}
{Bonato}, M., {Negrello}, M., {Cai}, Z.~Y., {et~al.} 2014, \mnras, 438, 2547,
  \dodoi{10.1093/mnras/stt2375}

\bibitem[{{Bonato} {et~al.}(2017){Bonato}, {Negrello}, {Mancuso}, {De Zotti},
  {Ciliegi}, {Cai}, {Lapi}, {Massardi}, {Bonaldi}, \& {Sajina}}]{Bonato2017}
{Bonato}, M., {Negrello}, M., {Mancuso}, C., {et~al.} 2017, \mnras, 469, 1912,
  \dodoi{10.1093/mnras/stx974}

\bibitem[{{Bonato} {et~al.}(2019){Bonato}, {De Zotti}, {Leisawitz}, {Negrello},
  {Massardi}, {Baronchelli}, {Cai}, {Bradford}, {Pope}, \&
  {Murphy}}]{Bonato2019}
{Bonato}, M., {De Zotti}, G., {Leisawitz}, D., {et~al.} 2019, \pasa, 36, e017,
  \dodoi{10.1017/pasa.2019.8}

\bibitem[{{Brown} \& {Wilson}(2019)}]{BrownWilson2019}
{Brown}, T., \& {Wilson}, C.~D. 2019, \apj, 879, 17,
  \dodoi{10.3847/1538-4357/ab2246}

\bibitem[{{Cai} {et~al.}(2014){Cai}, {Lapi}, {Bressan}, {De Zotti}, {Negrello},
  \& {Danese}}]{Cai2014}
{Cai}, Z.-Y., {Lapi}, A., {Bressan}, A., {et~al.} 2014, \apj, 785, 65,
  \dodoi{10.1088/0004-637X/785/1/65}

\bibitem[{{Cai} {et~al.}(2013){Cai}, {Lapi}, {Xia}, {De Zotti}, {Negrello},
  {Gruppioni}, {Rigby}, {Castex}, {Delabrouille}, \& {Danese}}]{Cai2013}
{Cai}, Z.-Y., {Lapi}, A., {Xia}, J.-Q., {et~al.} 2013, \apj, 768, 21,
  \dodoi{10.1088/0004-637X/768/1/21}

\bibitem[{{Carniani} {et~al.}(2015){Carniani}, {Maiolino}, {De Zotti},
  {Negrello}, {Marconi}, {Bothwell}, {Capak}, {Carilli}, {Castellano}, \&
  {Cristiani}}]{Carniani2015}
{Carniani}, S., {Maiolino}, R., {De Zotti}, G., {et~al.} 2015, \aap, 584, A78,
  \dodoi{10.1051/0004-6361/201525780}

\bibitem[{{Casey} {et~al.}(2014){Casey}, {Narayanan}, \& {Cooray}}]{Casey2014}
{Casey}, C.~M., {Narayanan}, D., \& {Cooray}, A. 2014, \physrep, 541, 45,
  \dodoi{10.1016/j.physrep.2014.02.009}

\bibitem[{{Chabrier}(2003)}]{Chabrier2003}
{Chabrier}, G. 2003, \pasp, 115, 763, \dodoi{10.1086/376392}

\bibitem[{{Chiosi} {et~al.}(1998){Chiosi}, {Bressan}, {Portinari}, \&
  {Tantalo}}]{Chiosi1998}
{Chiosi}, C., {Bressan}, A., {Portinari}, L., \& {Tantalo}, R. 1998, \aap, 339,
  355.
\newblock \doarXiv{astro-ph/9708123}

\bibitem[{{Chomiuk} \& {Povich}(2011)}]{ChomiukPovich2011}
{Chomiuk}, L., \& {Povich}, M.~S. 2011, \aj, 142, 197,
  \dodoi{10.1088/0004-6256/142/6/197}

\bibitem[{{Coppin} {et~al.}(2006){Coppin}, {Chapin}, {Mortier}, {Scott},
  {Borys}, {Dunlop}, {Halpern}, {Hughes}, {Pope}, \& {Scott}}]{Coppin2006}
{Coppin}, K., {Chapin}, E.~L., {Mortier}, A.~M.~J., {et~al.} 2006, \mnras, 372,
  1621, \dodoi{10.1111/j.1365-2966.2006.10961.x}

\bibitem[{{Cowley} {et~al.}(2019){Cowley}, {Lacey}, {Baugh}, {Cole}, {Frenk},
  \& {Lagos}}]{Cowley2019}
{Cowley}, W.~I., {Lacey}, C.~G., {Baugh}, C.~M., {et~al.} 2019, \mnras, 487,
  3082, \dodoi{10.1093/mnras/stz1398}

\bibitem[{{Davies} {et~al.}(2010){Davies}, {Baes}, {Bendo}, {Bianchi},
  {Bomans}, {Boselli}, {Clemens}, {Corbelli}, {Cortese}, {Dariush}, {De Looze},
  {di Serego Alighieri}, {Fadda}, {Fritz}, {Garcia-Appadoo}, {Gavazzi},
  {Giovanardi}, {Grossi}, {Hughes}, {Hunt}, {Jones}, {Madden}, {Pierini},
  {Pohlen}, {Sabatini}, {Smith}, {Verstappen}, {Vlahakis}, {Xilouris}, \&
  {Zibetti}}]{Davies2010}
{Davies}, J.~I., {Baes}, M., {Bendo}, G.~J., {et~al.} 2010, \aap, 518, L48,
  \dodoi{10.1051/0004-6361/201014571}

\bibitem[{{De Zotti} {et~al.}(2019){De Zotti}, {Bonato}, {Negrello},
  {Trombetti}, {Burigana}, {Herranz}, {L{\'o}pez-Caniego}, {Cai}, {Bonavera},
  \& {Gonz{\'a}lez-Nuevo}}]{DeZotti2019}
{De Zotti}, G., {Bonato}, M., {Negrello}, M., {et~al.} 2019, Frontiers in
  Astronomy and Space Sciences, 6, 53, \dodoi{10.3389/fspas.2019.00053}

\bibitem[{{Donevski} {et~al.}(2018){Donevski}, {Buat}, {Boone}, {Pappalardo},
  {Bethermin}, {Schreiber}, {Mazyed}, {Alvarez-Marquez}, \&
  {Duivenvoorden}}]{Donevski2018}
{Donevski}, D., {Buat}, V., {Boone}, F., {et~al.} 2018, \aap, 614, A33,
  \dodoi{10.1051/0004-6361/201731888}

\bibitem[{{Dowell} {et~al.}(2014){Dowell}, {Conley}, {Glenn}, {Arumugam},
  {Asboth}, {Aussel}, {Bertoldi}, {B{\'e}thermin}, {Bock}, \&
  {Boselli}}]{Dowell2014}
{Dowell}, C.~D., {Conley}, A., {Glenn}, J., {et~al.} 2014, \apj, 780, 75,
  \dodoi{10.1088/0004-637X/780/1/75}

\bibitem[{{Duivenvoorden} {et~al.}(2018){Duivenvoorden}, {Oliver}, {Scudder},
  {Greenslade}, {Riechers}, {Wilkins}, {Buat}, {Chapman}, {Clements}, \&
  {Cooray}}]{Duivenvoorden2018}
{Duivenvoorden}, S., {Oliver}, S., {Scudder}, J.~M., {et~al.} 2018, \mnras,
  477, 1099, \dodoi{10.1093/mnras/sty691}

\bibitem[{{Eales} {et~al.}(2010){Eales}, {Dunne}, {Clements}, {Cooray}, {De
  Zotti}, {Dye}, {Ivison}, {Jarvis}, {Lagache}, {Maddox}, {Negrello},
  {Serjeant}, {Thompson}, {Van Kampen}, {Amblard}, {Andreani}, {Baes},
  {Beelen}, {Bendo}, {Benford}, {Bertoldi}, {Bock}, {Bonfield}, {Boselli},
  {Bridge}, {Buat}, {Burgarella}, {Carlberg}, {Cava}, {Chanial}, {Charlot},
  {Christopher}, {Coles}, {Cortese}, {Dariush}, {da Cunha}, {Dalton}, {Danese},
  {Dannerbauer}, {Driver}, {Dunlop}, {Fan}, {Farrah}, {Frayer}, {Frenk},
  {Geach}, {Gardner}, {Gomez}, {Gonz{\'a}lez-Nuevo}, {Gonz{\'a}lez-Solares},
  {Griffin}, {Hardcastle}, {Hatziminaoglou}, {Herranz}, {Hughes}, {Ibar},
  {Jeong}, {Lacey}, {Lapi}, {Lawrence}, {Lee}, {Leeuw}, {Liske},
  {L{\'o}pez-Caniego}, {M{\"u}ller}, {Nandra}, {Panuzzo}, {Papageorgiou},
  {Patanchon}, {Peacock}, {Pearson}, {Phillipps}, {Pohlen}, {Popescu},
  {Rawlings}, {Rigby}, {Rigopoulou}, {Robotham}, {Rodighiero}, {Sansom},
  {Schulz}, {Scott}, {Smith}, {Sibthorpe}, {Smail}, {Stevens}, {Sutherland},
  {Takeuchi}, {Tedds}, {Temi}, {Tuffs}, {Trichas}, {Vaccari}, {Valtchanov},
  {van der Werf}, {Verma}, {Vieria}, {Vlahakis}, \& {White}}]{Eales2010}
{Eales}, S., {Dunne}, L., {Clements}, D., {et~al.} 2010, \pasp, 122, 499,
  \dodoi{10.1086/653086}

\bibitem[{{Eddington}(1913)}]{Eddington1913}
{Eddington}, A.~S. 1913, \mnras, 73, 359, \dodoi{10.1093/mnras/73.5.359}

\bibitem[{{Fudamoto} {et~al.}(2017){Fudamoto}, {Ivison}, {Oteo}, {Krips},
  {Zhang}, {Weiss}, {Dannerbauer}, {Omont}, {Chapman}, {Christensen},
  {Arumugam}, {Bertoldi}, {Bremer}, {Clements}, {Dunne}, {Eales}, {Greenslade},
  {Maddox}, {Martinez-Navajas}, {Michalowski}, {P{\'e}rez-Fournon}, {Riechers},
  {Simpson}, {Stalder}, {Valiante}, \& {van der Werf}}]{Fudamoto2017}
{Fudamoto}, Y., {Ivison}, R.~J., {Oteo}, I., {et~al.} 2017, \mnras, 472, 2028,
  \dodoi{10.1093/mnras/stx1956}

\bibitem[{{Geach} {et~al.}(2017){Geach}, {Dunlop}, {Halpern}, {Smail}, {van der
  Werf}, {Alexander}, {Almaini}, {Aretxaga}, {Arumugam}, \&
  {Asboth}}]{Geach2017}
{Geach}, J.~E., {Dunlop}, J.~S., {Halpern}, M., {et~al.} 2017, \mnras, 465,
  1789, \dodoi{10.1093/mnras/stw2721}

\bibitem[{{Gonz{\'a}lez-Nuevo} {et~al.}(2012){Gonz{\'a}lez-Nuevo}, {Lapi},
  {Fleuren}, {Bressan}, {Danese}, {De Zotti}, {Negrello}, {Cai}, {Fan},
  {Sutherland}, {Baes}, {Baker}, {Clements}, {Cooray}, {Dannerbauer}, {Dunne},
  {Dye}, {Eales}, {Frayer}, {Harris}, {Ivison}, {Jarvis}, {Micha{\l}owski},
  {L{\'o}pez-Caniego}, {Rodighiero}, {Rowlands}, {Serjeant}, {Scott}, {van der
  Werf}, {Auld}, {Buttiglione}, {Cava}, {Dariush}, {Fritz}, {Hopwood}, {Ibar},
  {Maddox}, {Pascale}, {Pohlen}, {Rigby}, {Smith}, \&
  {Temi}}]{GonzalezNuevo2012}
{Gonz{\'a}lez-Nuevo}, J., {Lapi}, A., {Fleuren}, S., {et~al.} 2012, \apj, 749,
  65, \dodoi{10.1088/0004-637X/749/1/65}

\bibitem[{{Gralla} {et~al.}(2019){Gralla}, {Marriage}, {Addison}, {Baker},
  {Bond}, {Crichton}, {Datta}, {Devlin}, {Dunkley}, \&
  {D{\"u}nner}}]{Gralla2019}
{Gralla}, M.~B., {Marriage}, T.~A., {Addison}, G., {et~al.} 2019, arXiv
  e-prints, arXiv:1905.04592.
\newblock \doarXiv{1905.04592}

\bibitem[{{Granato} \& {Danese}(1994)}]{GranatoDanese1994}
{Granato}, G.~L., \& {Danese}, L. 1994, \mnras, 268, 235,
  \dodoi{10.1093/mnras/268.1.235}

\bibitem[{{Granato} {et~al.}(2004){Granato}, {De Zotti}, {Silva}, {Bressan}, \&
  {Danese}}]{Granato2004}
{Granato}, G.~L., {De Zotti}, G., {Silva}, L., {Bressan}, A., \& {Danese}, L.
  2004, \apj, 600, 580, \dodoi{10.1086/379875}

\bibitem[{{Gruppioni} {et~al.}(2013){Gruppioni}, {Pozzi}, {Rodighiero},
  {Delvecchio}, {Berta}, {Pozzetti}, {Zamorani}, {Andreani}, {Cimatti}, \&
  {Ilbert}}]{Gruppioni2013}
{Gruppioni}, C., {Pozzi}, F., {Rodighiero}, G., {et~al.} 2013, \mnras, 432, 23,
  \dodoi{10.1093/mnras/stt308}

\bibitem[{{Gruppioni} {et~al.}(2015){Gruppioni}, {Calura}, {Pozzi},
  {Delvecchio}, {Berta}, {De Lucia}, {Fontanot}, {Franceschini}, {Marchetti},
  \& {Menci}}]{Gruppioni2015}
{Gruppioni}, C., {Calura}, F., {Pozzi}, F., {et~al.} 2015, \mnras, 451, 3419,
  \dodoi{10.1093/mnras/stv1204}

\bibitem[{{Hatsukade} {et~al.}(2018){Hatsukade}, {Kohno}, {Yamaguchi},
  {Umehata}, {Ao}, {Aretxaga}, {Caputi}, {Dunlop}, {Egami}, {Espada},
  {Fujimoto}, {Hayatsu}, {Hughes}, {Ikarashi}, {Iono}, {Ivison}, {Kawabe},
  {Kodama}, {Lee}, {Matsuda}, {Nakanishi}, {Ohta}, {Ouchi}, {Rujopakarn},
  {Suzuki}, {Tamura}, {Ueda}, {Wang}, {Wang}, {Wilson}, {Yoshimura}, \&
  {Yun}}]{Hatsukade2018}
{Hatsukade}, B., {Kohno}, K., {Yamaguchi}, Y., {et~al.} 2018, \pasj, 70, 105,
  \dodoi{10.1093/pasj/psy104}

\bibitem[{{Hogg} \& {Turner}(1998)}]{HoggTurner1998}
{Hogg}, D.~W., \& {Turner}, E.~L. 1998, \pasp, 110, 727, \dodoi{10.1086/316173}

\bibitem[{{Hughes} {et~al.}(1998){Hughes}, {Serjeant}, {Dunlop},
  {Rowan-Robinson}, {Blain}, {Mann}, {Ivison}, {Peacock}, {Efstathiou}, \&
  {Gear}}]{Hughes1998}
{Hughes}, D.~H., {Serjeant}, S., {Dunlop}, J., {et~al.} 1998, \nat, 394, 241,
  \dodoi{10.1038/28328}

\bibitem[{{Ivison} {et~al.}(2016){Ivison}, {Lewis}, {Weiss}, {Arumugam},
  {Simpson}, {Holland}, {Maddox}, {Dunne}, {Valiante}, \& {van der
  Werf}}]{Ivison2016}
{Ivison}, R.~J., {Lewis}, A.~J.~R., {Weiss}, A., {et~al.} 2016, \apj, 832, 78,
  \dodoi{10.3847/0004-637X/832/1/78}

\bibitem[{{Jin} {et~al.}(2018){Jin}, {Daddi}, {Liu}, {Smol{\v{c}}i{\'c}},
  {Schinnerer}, {Calabr{\`o}}, {Gu}, {Delhaize}, {Delvecchio}, \&
  {Gao}}]{Jin2018}
{Jin}, S., {Daddi}, E., {Liu}, D., {et~al.} 2018, \apj, 864, 56,
  \dodoi{10.3847/1538-4357/aad4af}

\bibitem[{{Kaviani} {et~al.}(2003){Kaviani}, {Haehnelt}, \&
  {Kauffmann}}]{Kaviani2003}
{Kaviani}, A., {Haehnelt}, M.~G., \& {Kauffmann}, G. 2003, \mnras, 340, 739,
  \dodoi{10.1046/j.1365-8711.2003.06318.x}

\bibitem[{{Kennicutt}(1998)}]{Kennicutt1998}
{Kennicutt}, Robert~C., J. 1998, \araa, 36, 189,
  \dodoi{10.1146/annurev.astro.36.1.189}

\bibitem[{{Kennicutt} \& {Evans}(2012)}]{KennicuttEvans2012}
{Kennicutt}, R.~C., \& {Evans}, N.~J. 2012, \araa, 50, 531,
  \dodoi{10.1146/annurev-astro-081811-125610}

\bibitem[{{Kroupa} {et~al.}(2013){Kroupa}, {Weidner}, {Pflamm-Altenburg},
  {Thies}, {Dabringhausen}, {Marks}, \& {Maschberger}}]{Kroupa2013}
{Kroupa}, P., {Weidner}, C., {Pflamm-Altenburg}, J., {et~al.} 2013, {The
  Stellar and Sub-Stellar Initial Mass Function of Simple and Composite
  Populations}, ed. T.~D. {Oswalt} \& G.~{Gilmore}, Vol.~5, 115

\bibitem[{{Lacey} {et~al.}(2016){Lacey}, {Baugh}, {Frenk}, {Benson}, {Bower},
  {Cole}, {Gonzalez-Perez}, {Helly}, {Lagos}, \& {Mitchell}}]{Lacey2016}
{Lacey}, C.~G., {Baugh}, C.~M., {Frenk}, C.~S., {et~al.} 2016, \mnras, 462,
  3854, \dodoi{10.1093/mnras/stw1888}

\bibitem[{{Lagos} {et~al.}(2018){Lagos}, {Tobar}, {Robotham}, {Obreschkow},
  {Mitchell}, {Power}, \& {Elahi}}]{Lagos2018}
{Lagos}, C. d.~P., {Tobar}, R.~J., {Robotham}, A. S.~G., {et~al.} 2018, \mnras,
  481, 3573, \dodoi{10.1093/mnras/sty2440}

\bibitem[{{Lagos} {et~al.}(2019){Lagos}, {Robotham}, {Trayford}, {Tobar},
  {Bravo}, {Bellstedt}, {Davies}, {Driver}, {Elahi}, {Obreschkow}, \&
  {Power}}]{Lagos2019}
{Lagos}, C. d.~P., {Robotham}, A. S.~G., {Trayford}, J.~W., {et~al.} 2019,
  \mnras, 489, 4196, \dodoi{10.1093/mnras/stz2427}

\bibitem[{{Lapi} {et~al.}(2006){Lapi}, {Shankar}, {Mao}, {Granato}, {Silva},
  {De Zotti}, \& {Danese}}]{Lapi2006}
{Lapi}, A., {Shankar}, F., {Mao}, J., {et~al.} 2006, \apj, 650, 42,
  \dodoi{10.1086/507122}

\bibitem[{{Lapi} {et~al.}(2011){Lapi}, {Gonz{\'a}lez-Nuevo}, {Fan}, {Bressan},
  {De Zotti}, {Danese}, {Negrello}, {Dunne}, {Eales}, \& {Maddox}}]{Lapi2011}
{Lapi}, A., {Gonz{\'a}lez-Nuevo}, J., {Fan}, L., {et~al.} 2011, \apj, 742, 24,
  \dodoi{10.1088/0004-637X/742/1/24}

\bibitem[{{Leisawitz} {et~al.}(2018){Leisawitz}, {Amatucci}, {Carter},
  {DiPirro}, {Flores}, {Staguhn}, {Wu}, {Allen}, {Arenberg}, {Armus},
  {Battersby}, {Bauer}, {Bell}, {Beltran}, {Benford}, {Bergin}, {Bradford},
  {Bradley}, {Burgarella}, {Carey}, {Chi}, {Cooray}, {Corsetti}, {De Beck},
  {Denis}, {Dewell}, {East}, {Edgington}, {Ennico}, {Fantano}, {Feller},
  {Folta}, {Fortney}, {Generie}, {Gerin}, {Granger}, {Harpole}, {Harvey},
  {Helmich}, {Hilliard}, {Howard}, {Jacoby}, {Jamil}, {Kataria}, {Knight},
  {Knollenberg}, {Lightsey}, {Lipscy}, {Mamajek}, {Martins}, {Meixner},
  {Melnick}, {Milam}, {Mooney}, {Moseley}, {Narayanan}, {Neff}, {Nguyen},
  {Nordt}, {Olson}, {Padgett}, {Petach}, {Petro}, {Pohner}, {Pontoppidan},
  {Pope}, {Ramspacher}, {Roellig}, {Sakon}, {Sandin}, {Sandstrom}, {Scott},
  {Sheth}, {Steeves}, {Stevenson}, {Stokowski}, {Stoneking}, {Su}, {Tajdaran},
  {Tompkins}, {Vieira}, {Webster}, {Wiedner}, {Wright}, \&
  {Zmuidzinas}}]{Leisawitz2018}
{Leisawitz}, D., {Amatucci}, E., {Carter}, R., {et~al.} 2018, in Society of
  Photo-Optical Instrumentation Engineers (SPIE) Conference Series, Vol. 10698,
  \procspie, 1069815

\bibitem[{{Liang} {et~al.}(2019){Liang}, {Feldmann}, {Kere{\v{s}}}, {Scoville},
  {Hayward}, {Faucher-Gigu{\`e}re}, {Schreiber}, {Ma}, {Hopkins}, \&
  {Quataert}}]{Liang2019}
{Liang}, L., {Feldmann}, R., {Kere{\v{s}}}, D., {et~al.} 2019, \mnras, 489,
  1397, \dodoi{10.1093/mnras/stz2134}

\bibitem[{{Madau} {et~al.}(1998){Madau}, {Pozzetti}, \&
  {Dickinson}}]{Madau1998}
{Madau}, P., {Pozzetti}, L., \& {Dickinson}, M. 1998, \apj, 498, 106,
  \dodoi{10.1086/305523}

\bibitem[{{Magnelli} {et~al.}(2013){Magnelli}, {Popesso}, {Berta}, {Pozzi},
  {Elbaz}, {Lutz}, {Dickinson}, {Altieri}, {Andreani}, {Aussel},
  {B{\'e}thermin}, {Bongiovanni}, {Cepa}, {Charmandaris}, {Chary}, {Cimatti},
  {Daddi}, {F{\"o}rster Schreiber}, {Genzel}, {Gruppioni}, {Harwit}, {Hwang},
  {Ivison}, {Magdis}, {Maiolino}, {Murphy}, {Nordon}, {Pannella}, {P{\'e}rez
  Garc{\'\i}a}, {Poglitsch}, {Rosario}, {Sanchez-Portal}, {Santini}, {Scott},
  {Sturm}, {Tacconi}, \& {Valtchanov}}]{Magnelli2013}
{Magnelli}, B., {Popesso}, P., {Berta}, S., {et~al.} 2013, \aap, 553, A132,
  \dodoi{10.1051/0004-6361/201321371}

\bibitem[{{Magnelli} {et~al.}(2019){Magnelli}, {Karim}, {Staguhn},
  {Kov{\'a}cs}, {Jim{\'e}nez-Andrade}, {Casey}, {Zavala}, {Schinnerer},
  {Sargent}, \& {Aravena}}]{Magnelli2019}
{Magnelli}, B., {Karim}, A., {Staguhn}, J., {et~al.} 2019, \apj, 877, 45,
  \dodoi{10.3847/1538-4357/ab1912}

\bibitem[{{Niemi} {et~al.}(2012){Niemi}, {Somerville}, {Ferguson}, {Huang},
  {Lotz}, \& {Koekemoer}}]{Niemi2012}
{Niemi}, S.-M., {Somerville}, R.~S., {Ferguson}, H.~C., {et~al.} 2012, \mnras,
  421, 1539, \dodoi{10.1111/j.1365-2966.2012.20425.x}

\bibitem[{{Oliver} {et~al.}(2012){Oliver}, {Bock}, {Altieri}, {Amblard},
  {Arumugam}, {Aussel}, {Babbedge}, {Beelen}, {B{\'e}thermin}, {Blain},
  {Boselli}, {Bridge}, {Brisbin}, {Buat}, {Burgarella},
  {Castro-Rodr{\'\i}guez}, {Cava}, {Chanial}, {Cirasuolo}, {Clements},
  {Conley}, {Conversi}, {Cooray}, {Dowell}, {Dubois}, {Dwek}, {Dye}, {Eales},
  {Elbaz}, {Farrah}, {Feltre}, {Ferrero}, {Fiolet}, {Fox}, {Franceschini},
  {Gear}, {Giovannoli}, {Glenn}, {Gong}, {Gonz{\'a}lez Solares}, {Griffin},
  {Halpern}, {Harwit}, {Hatziminaoglou}, {Heinis}, {Hurley}, {Hwang}, {Hyde},
  {Ibar}, {Ilbert}, {Isaak}, {Ivison}, {Lagache}, {Le Floc'h}, {Levenson},
  {Faro}, {Lu}, {Madden}, {Maffei}, {Magdis}, {Mainetti}, {Marchetti},
  {Marsden}, {Marshall}, {Mortier}, {Nguyen}, {O'Halloran}, {Omont}, {Page},
  {Panuzzo}, {Papageorgiou}, {Patel}, {Pearson}, {P{\'e}rez-Fournon}, {Pohlen},
  {Rawlings}, {Raymond}, {Rigopoulou}, {Riguccini}, {Rizzo}, {Rodighiero},
  {Roseboom}, {Rowan-Robinson}, {S{\'a}nchez Portal}, {Schulz}, {Scott},
  {Seymour}, {Shupe}, {Smith}, {Stevens}, {Symeonidis}, {Trichas}, {Tugwell},
  {Vaccari}, {Valtchanov}, {Vieira}, {Viero}, {Vigroux}, {Wang}, {Ward},
  {Wardlow}, {Wright}, {Xu}, \& {Zemcov}}]{Oliver2012}
{Oliver}, S.~J., {Bock}, J., {Altieri}, B., {et~al.} 2012, \mnras, 424, 1614,
  \dodoi{10.1111/j.1365-2966.2012.20912.x}

\bibitem[{{Padoan} {et~al.}(1997){Padoan}, {Nordlund}, \& {Jones}}]{Padoan1997}
{Padoan}, P., {Nordlund}, A., \& {Jones}, B. J.~T. 1997, \mnras, 288, 145,
  \dodoi{10.1093/mnras/288.1.145}

\bibitem[{{Papadopoulos} {et~al.}(2011){Papadopoulos}, {Thi}, {Miniati}, \&
  {Viti}}]{Papadopoulos2011}
{Papadopoulos}, P.~P., {Thi}, W.-F., {Miniati}, F., \& {Viti}, S. 2011, \mnras,
  414, 1705, \dodoi{10.1111/j.1365-2966.2011.18504.x}

\bibitem[{{Pillepich} {et~al.}(2015){Pillepich}, {Madau}, \&
  {Mayer}}]{Pillepich2015}
{Pillepich}, A., {Madau}, P., \& {Mayer}, L. 2015, \apj, 799, 184,
  \dodoi{10.1088/0004-637X/799/2/184}

\bibitem[{{Planck Collaboration VI}(2018)}]{PlanckCollaboration2018parameters}
{Planck Collaboration VI}. 2018, arXiv e-prints, arXiv:1807.06209.
\newblock \doarXiv{1807.06209}

\bibitem[{{Richards} {et~al.}(2006){Richards}, {Lacy}, {Storrie-Lombardi},
  {Hall}, {Gallagher}, {Hines}, {Fan}, {Papovich}, {Vanden Berk}, {Trammell},
  {Schneider}, {Vestergaard}, {York}, {Jester}, {Anderson}, {Budav{\'a}ri}, \&
  {Szalay}}]{Richards2006}
{Richards}, G.~T., {Lacy}, M., {Storrie-Lombardi}, L.~J., {et~al.} 2006, \apjs,
  166, 470, \dodoi{10.1086/506525}

\bibitem[{{Riechers} {et~al.}(2013){Riechers}, {Bradford}, {Clements},
  {Dowell}, {P{\'e}rez-Fournon}, {Ivison}, {Bridge}, {Conley}, {Fu}, {Vieira},
  {Wardlow}, {Calanog}, {Cooray}, {Hurley}, {Neri}, {Kamenetzky}, {Aguirre},
  {Altieri}, {Arumugam}, {Benford}, {B{\'e}thermin}, {Bock}, {Burgarella},
  {Cabrera-Lavers}, {Chapman}, {Cox}, {Dunlop}, {Earle}, {Farrah}, {Ferrero},
  {Franceschini}, {Gavazzi}, {Glenn}, {Solares}, {Gurwell}, {Halpern},
  {Hatziminaoglou}, {Hyde}, {Ibar}, {Kov{\'a}cs}, {Krips}, {Lupu}, {Maloney},
  {Martinez-Navajas}, {Matsuhara}, {Murphy}, {Naylor}, {Nguyen}, {Oliver},
  {Omont}, {Page}, {Petitpas}, {Rangwala}, {Roseboom}, {Scott}, {Smith},
  {Staguhn}, {Streblyanska}, {Thomson}, {Valtchanov}, {Viero}, {Wang},
  {Zemcov}, \& {Zmuidzinas}}]{Riechers2013}
{Riechers}, D.~A., {Bradford}, C.~M., {Clements}, D.~L., {et~al.} 2013, \nat,
  496, 329, \dodoi{10.1038/nature12050}

\bibitem[{{Schreiber} {et~al.}(2018){Schreiber}, {Elbaz}, {Pannella}, {Ciesla},
  {Wang}, \& {Franco}}]{Schreiber2018}
{Schreiber}, C., {Elbaz}, D., {Pannella}, M., {et~al.} 2018, \aap, 609, A30,
  \dodoi{10.1051/0004-6361/201731506}

\bibitem[{{Schreiber} {et~al.}(2016){Schreiber}, {Elbaz}, {Pannella}, {Ciesla},
  {Wang}, {Koekemoer}, {Rafelski}, \& {Daddi}}]{Schreiber2016}
---. 2016, \aap, 589, A35, \dodoi{10.1051/0004-6361/201527200}

\bibitem[{{Sheth} \& {Tormen}(1999)}]{ShethTormen1999}
{Sheth}, R.~K., \& {Tormen}, G. 1999, \mnras, 308, 119,
  \dodoi{10.1046/j.1365-8711.1999.02692.x}

\bibitem[{{Silk} \& {Mamon}(2012)}]{SilkMamon2012}
{Silk}, J., \& {Mamon}, G.~A. 2012, Research in Astronomy and Astrophysics, 12,
  917, \dodoi{10.1088/1674-4527/12/8/004}

\bibitem[{{Silva} {et~al.}(1998){Silva}, {Granato}, {Bressan}, \&
  {Danese}}]{Silva1998}
{Silva}, L., {Granato}, G.~L., {Bressan}, A., \& {Danese}, L. 1998, \apj, 509,
  103, \dodoi{10.1086/306476}

\bibitem[{{Sliwa} {et~al.}(2017){Sliwa}, {Wilson}, {Aalto}, \&
  {Privon}}]{Sliwa2017}
{Sliwa}, K., {Wilson}, C.~D., {Aalto}, S., \& {Privon}, G.~C. 2017, \apjl, 840,
  L11, \dodoi{10.3847/2041-8213/aa6ea4}

\bibitem[{{Smail} {et~al.}(1997){Smail}, {Ivison}, \& {Blain}}]{Smail1997}
{Smail}, I., {Ivison}, R.~J., \& {Blain}, A.~W. 1997, \apjl, 490, L5,
  \dodoi{10.1086/311017}

\bibitem[{{Stacey} {et~al.}(2018){Stacey}, {Aravena}, {Basu}, {Battaglia},
  {Beringue}, {Bertoldi}, {Bond}, {Breysse}, {Bustos}, {Chapman}, {Chung},
  {Cothard}, {Erler}, {Fich}, {Foreman}, {Gallardo}, {Giovanelli}, {Graf},
  {Haynes}, {Herrera-Camus}, {Herter}, {Hlo{\v{z}}ek}, {Johnstone}, {Keating},
  {Magnelli}, {Meerburg}, {Meyers}, {Murray}, {Niemack}, {Nikola}, {Nolta},
  {Parshley}, {Riechers}, {Schilke}, {Scott}, {Stein}, {Stevens}, {Stutzki},
  {Vavagiakis}, \& {Viero}}]{Stacey2018}
{Stacey}, G.~J., {Aravena}, M., {Basu}, K., {et~al.} 2018, in Society of
  Photo-Optical Instrumentation Engineers (SPIE) Conference Series, Vol. 10700,
  \procspie, 107001M

\bibitem[{{Swinbank} {et~al.}(2010){Swinbank}, {Smail}, {Longmore}, {Harris},
  {Baker}, {De Breuck}, {Richard}, {Edge}, {Ivison}, {Blundell}, {Coppin},
  {Cox}, {Gurwell}, {Hainline}, {Krips}, {Lundgren}, {Neri}, {Siana},
  {Siringo}, {Stark}, {Wilner}, \& {Younger}}]{Swinbank2010}
{Swinbank}, A.~M., {Smail}, I., {Longmore}, S., {et~al.} 2010, \nat, 464, 733,
  \dodoi{10.1038/nature08880}

\bibitem[{{Thomas} {et~al.}(2010){Thomas}, {Maraston}, {Schawinski}, {Sarzi},
  \& {Silk}}]{Thomas2010}
{Thomas}, D., {Maraston}, C., {Schawinski}, K., {Sarzi}, M., \& {Silk}, J.
  2010, \mnras, 404, 1775, \dodoi{10.1111/j.1365-2966.2010.16427.x}

\bibitem[{{Wang} {et~al.}(2011){Wang}, {Navarro}, {Frenk}, {White}, {Springel},
  {Jenkins}, {Helmi}, {Ludlow}, \& {Vogelsberger}}]{Wang2011}
{Wang}, J., {Navarro}, J.~F., {Frenk}, C.~S., {et~al.} 2011, \mnras, 413, 1373,
  \dodoi{10.1111/j.1365-2966.2011.18220.x}

\bibitem[{{Wang} {et~al.}(2019){Wang}, {Pearson}, {Cowley}, {Trayford},
  {B{\'e}thermin}, {Gruppioni}, {Hurley}, \& {Micha{\l}owski}}]{Wang2019}
{Wang}, L., {Pearson}, W.~J., {Cowley}, W., {et~al.} 2019, \aap, 624, A98,
  \dodoi{10.1051/0004-6361/201834093}

\bibitem[{{Xia} {et~al.}(2012){Xia}, {Negrello}, {Lapi}, {De Zotti}, {Danese},
  \& {Viel}}]{Xia2012}
{Xia}, J.-Q., {Negrello}, M., {Lapi}, A., {et~al.} 2012, \mnras, 422, 1324,
  \dodoi{10.1111/j.1365-2966.2012.20705.x}

\bibitem[{{Zavala} {et~al.}(2018){Zavala}, {Monta{\~n}a}, {Hughes}, {Yun},
  {Ivison}, {Valiante}, {Wilner}, {Spilker}, {Aretxaga}, {Eales},
  {Avila-Reese}, {Ch{\'a}vez}, {Cooray}, {Dannerbauer}, {Dunlop}, {Dunne},
  {G{\'o}mez-Ruiz}, {Micha{\l}owski}, {Narayanan}, {Nayyeri}, {Oteo}, {Rosa
  Gonz{\'a}lez}, {S{\'a}nchez-Arg{\"u}elles}, {Schloerb}, {Serjeant}, {Smith},
  {Terlevich}, {Vega}, {Villalba}, {van der Werf}, {Wilson}, \&
  {Zeballos}}]{Zavala2018}
{Zavala}, J.~A., {Monta{\~n}a}, A., {Hughes}, D.~H., {et~al.} 2018, Nature
  Astronomy, 2, 56, \dodoi{10.1038/s41550-017-0297-8}

\bibitem[{{Zhang} {et~al.}(2016){Zhang}, {Papadopoulos}, {Ivison}, {Galametz},
  {Smith}, \& {Xilouris}}]{Zhang2016}
{Zhang}, Z.-Y., {Papadopoulos}, P.~P., {Ivison}, R.~J., {et~al.} 2016, Royal
  Society Open Science, 3, 160025, \dodoi{10.1098/rsos.160025}

\bibitem[{{Zhang} {et~al.}(2018){Zhang}, {Romano}, {Ivison}, {Papadopoulos}, \&
  {Matteucci}}]{Zhang2018}
{Zhang}, Z.-Y., {Romano}, D., {Ivison}, R.~J., {Papadopoulos}, P.~P., \&
  {Matteucci}, F. 2018, \nat, 558, 260, \dodoi{10.1038/s41586-018-0196-x}

\end{thebibliography}

\end{document}